\documentclass[12pt]{article}
\usepackage{graphicx}
\usepackage{amsmath}
\usepackage{hyperref}
\hypersetup{
	colorlinks=true,
	citecolor=red,
	linkcolor=blue,
	filecolor=magenta,
	urlcolor=cyan
}
\usepackage{lineno}
\usepackage{xcolor}
\usepackage[symbol]{footmisc}
\usepackage{authblk}

\renewcommand{\thefootnote}{\fnsymbol{footnote}}

\newcommand\pubnumber{CIPANP2018-Quinn}
\newcommand\pubdate{\today}

\def\Title#1{\begin{center} {\Large #1 } \end{center}}
\def\Author#1{\begin{center}{ \sc #1} \end{center}}
\def\Address#1{\begin{center}{ \it #1} \end{center}}

\newcommand\pubblock{\rightline{\begin{tabular}{l} \pubnumber\\
         \pubdate  \end{tabular}}}
\newenvironment{Abstract}{\begin{quotation}  }{\end{quotation}}
\newenvironment{Presented}{\begin{quotation} \begin{center} 
             PRESENTED AT\end{center}\bigskip 
      \begin{center}\begin{large}}{\end{large}\end{center} \end{quotation}}
\def\Acknowledgements{\bigskip  \bigskip \begin{center} \begin{large}
             \bf ACKNOWLEDGEMENTS \end{large}\end{center}}

\def\beq{\begin{equation}}
\def\eeq#1{\label{#1}\end{equation}}
\def\eeqn{\end{equation}}

\def\beqa{\begin{eqnarray}}
\def\eeqa#1{\label{#1}\end{eqnarray}}
\def\eeqan{\end{eqnarray}}

\let\bar=\overbar

\def\Dslash{\not{\hbox{\kern-4pt $D$}}}
\def\dslash{\not{\hbox{\kern-2pt $\del$}}}

\def\msb{{\bar{\ssstyle M \kern -1pt S}}}

\begin{document}
\begin{titlepage}
\pubblock

\vfill
\Title{GAPS: A New Cosmic Ray Anti-matter Experiment}
\vfill
\Author{S.~Quinn$^{1}$\footnote[1]{spq@ucla.edu},
	T.~Aramaki$^{2}$, R.~Bird$^{1}$, M.~Boezio$^{3}$, 
	S.E.~Boggs$^{4}$, V.~Bonvicini$^{3}$, D.~Campana$^{5}$, W.W.~Craig$^{6}$,  
	P.~von~Doetinchem$^{7}$, E.~Everson$^{3}$,  L.~Fabris$^{8}$,  F.~Gahbauer$^{9}$, C.~Gerrity$^{7}$,
	H.~Fuke$^{10}$, C.J.~Hailey$^{9}$, T.~Hayashi$^{1}$, C.~Kato$^{11}$, A.~Kawachi$^{12}$,  M.~Kozai$^{10}$, A.~Lowell$^{4}$, 
	M.~Martucci$^{13}$,  S.I.~Mognet$^{14}$, R.~Munini$^{3}$,  K.~Munakata$^{11}$, S.~Okazaki$^{10}$, R.A.~Ong$^{1}$,  G.~Osteria$^{5}$, K.~Perez$^{15}$, J.~Ryan$^{1}$,  V.~Re$^{16}$, F.~Rogers$^{15}$, N.~Saffold$^{9}$, Y.~Shimizu$^{17}$, R.~Sparvoli$^{13}$, A.~Stoessl$^{7}$, E.~Vannuccini$^{18}$, A.~Yoshida$^{19}$, T.~Yoshida$^{10}$, G.~Zampa$^{3}$ and J.~Zweerink$^{1}$
}

\Address{$^{1}$University of California, Los Angeles, USA \\
	$^{2}$SLAC National Accelerator Laboratory, USA \\
	$^{3}$INFN, Sezione di Trieste, Trieste, Italy \\
	$^{4}$University of California, San Diego, USA \\
	$^{5}$INFN, Sezione di Napoli, Naples, Italy \\
	$^{6}$Lawrence Livermore National Laboratory, USA \\
	$^{7}$University of Hawaii at Manoa, USA \\
	$^{8}$Oak Ridge National Laboratory, USA \\
	$^{9}$Columbia University, USA \\
	$^{10}$Japan Aerospace Exploration Agency, Japan \\
	$^{11}$Shinshu University, Japan \\
	$^{12}$Tokai University, Japan \\
	$^{13}$INFN, Sezione di Roma ``Tor Vergata'', Rome, Italy \\
	$^{14}$Pennsylvania State University, USA \\
	$^{15}$Massachusetts Institute of Technology, USA \\
	$^{16}$Universit\`{a} di Bergamo, Bergamo, Italy \\
	$^{17}$Kanagawa University, Japan \\
	$^{18}$INFN, Sezione di Firenze, Florence, Italy \\
	$^{19}$Aoyama Gakuin University, Japan
}

\vfill


\begin{Abstract}
The General AntiParticle Spectrometer (GAPS) is a balloon-borne instrument designed to detect cosmic-ray antimatter using the novel exotic atom technique, obviating the strong magnetic fields required by experiments like AMS, PAMELA, or BESS. It will be sensitive to primary antideuterons with kinetic energies of $\approx0.05$--0.2 GeV/nucleon, providing some overlap with the previously mentioned experiments at the highest energies. For $3\times35$ day balloon flights, and standard classes of primary antideuteron propagation models, GAPS will be sensitive to $m_{\mathrm{DM}}\approx10$--100 GeV c$^{-2}$ WIMPs with a dark-matter flux to astrophysical flux ratio approaching 100. This clean primary channel is a key feature of GAPS and is crucial for a rare event search. Additionally, the antiproton spectrum will be extended with high statistics measurements to cover the $0.07 \leq E \leq 0.25 $ GeV domain. For $E>0.2$ GeV GAPS data will be complementary to existing experiments, while $E<0.2$ GeV explores a new regime. The first flight is scheduled for late 2020 in Antarctica. These proceedings will describe the astrophysical processes and backgrounds relevant to the dark matter search, a brief discussion of detector operation, and construction progress made to date.
\end{Abstract}
\vfill
\begin{Presented}
Thirteenth Conference on the Intersections of Particle and Nuclear Physics\\
Palm Springs, CA,  May 29 -- June 3, 2018
\end{Presented}
\vfill
\end{titlepage}
\def\thefootnote{\fnsymbol{footnote}}
\setcounter{footnote}{0}

\section{Introduction}
The fundamental nature of the majority of matter in the universe remains poorly understood. While the standard model provides an accurate framework for describing strong and electroweak interactions, it fails to predict a large portion of the gravitationally interacting matter observed in the universe. Cosmological data and the cold dark matter model ($\Lambda$CDM) suggest a non-baryonic matter density of $\Omega_c \approx 27$\% (using a physical non-baryon density parameter $\Omega_c h^2 = 0.1199$ with reduced Hubble constant $h=67.26 / 100$ km s$^{-1}$ Mpc$^{-1}$) \cite{Planck2015}. Supporting observational evidence includes galaxy rotation curves \cite{Rubin1970}, weak lensing observations \cite{Clowe2004} and deuterium to hydrogen abundance ratios \cite{OMeara2001}.

To explain the phenomenon one approach involves extending the standard model of particle physics to add a new particle (or particles) that could comprise the dark matter. There are a multitude of possibilities here, but a well motivated candidate is weakly interacting massive particles (WIMP). Some examples include right-handed or sterile sneutrinos, neutralinos, and gravitinos---all are the lightest supersymmetric partner (LSP) of their corresponding supersymmetric (SUSY) theory. Similar stable particles emerge from extra-dimensional theories, called the lightest Kaluza-Klein particle (LKP). The high level motivation for these theories is not necessarily due to their applicability to dark matter, but their usefulness in resolving other particle physics puzzles such as the hierarchy problem and/or gauge interaction unification. While certain WIMP models with masses from 5--200 GeV c$^{-2}$ have been ruled out by a variety of direct detection experiments (see \cite{DMSurvey2018}), candidates below 30 GeV c$^{-2}$ or above 200 GeV c$^{-2}$ remain \cite{Donato2018}. The aim of GAPS is to indirectly detect the decay or annihilation of these primary dark matter candidates using cosmic ray antimatter arriving in the upper atmosphere.

\subsection{Cosmic ray antimatter as probes for dark matter}
\begin{figure}
	\includegraphics[scale=0.38]{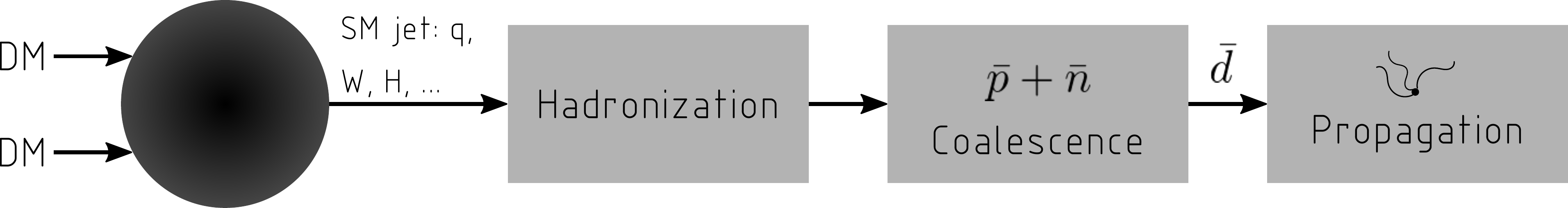}
	\caption{\label{fig:cartoon}High level diagram of primary dark matter interaction resulting in standard model fragments which then distill into hadrons. Anti-deuterons are then formed via a coalescence process \cite{Donato2008}, which eventually propagate to the solar system undergoing diffusion due to galactic magnetic fields.}
\end{figure}

Indirect dark matter detectors are designed to measure daughter standard model particles such as $e^{\pm}$, $\gamma$, $\bar{p}$, and $\bar{d}$ resulting from the decay or annihilation of a given dark matter species. A cartoon of the process for $\bar{d}$ production is shown in Figure \ref{fig:cartoon}.

The flux of cosmic ray antimatter at the top of the atmosphere (TOA) due to dark matter interactions, termed the ``primary flux'', depends on several factors: Galaxy density profile (e.g. NFW, Einasto, etc.), branching ratios of decay/annihilation channels, hadronization and coalescence models, and solar modulation. In addition to $\bar{d}$ primary flux, there is a ``secondary flux'' generated by ordinary spallation of high-energy protons in the interstellar medium. These particles are produced by ordinary galactic astrophysical processes such as shock acceleration in supernova remnants.

For antiprotons the kinetic energy peak due astrophysical spallation occurs at ~12 GeV, which is consistent with measurements of balloon and low Earth orbit experiments \cite{Boezio_97,Boezio_01,Asaoka_02,Abe_08,Abe_17,ams_16}. This secondary peak tends to be lower in kinetic energy (hundreds of MeV to a few GeV) from $\bar{p}/p$ peaks due to DM interactions. Although the separation is important, the flux contribution due to DM is not large, making it difficult to distinguish between a DM origin of antimatter, compared to natural background. For a sample of models, and the expected GAPS sensitivity, see \cite{aramaki_14}.

Anti-deuteron spectral peaks are well separated in terms of kinetic energy, but unlike antiprotons, the flux contribution due to DM production exceeds the background by up to a few orders of magnitude (depending on what propagation model is considered) below the secondary peak \cite{aramaki_16}. This makes $\bar{d}$ a promising channel for indirect DM detection---a half dozen events would be highly significant. Although it is a relatively clean channel, $\bar{d}$ measurements demand high flux sensitivity for $T<1$ GeV/$n$ primaries.

\subsection{Survey of low energy cosmic ray observations}
Presently there is one instrument capable of detecting $\bar{d}$ with kinetic energies between 100 MeV and 10 GeV:  AMS-02. It has similar sensitivity to GAPS, but does not extend below kinetic energies of ~200 MeV due to bore size, magnetic field strength limitations and geomagnetic cutoff. BESS has similar kinetic energy coverage, but a sensitivity cutoff that is roughly $100\times$ worse than AMS-02 and GAPS---primarily due to it's smaller acceptance and duty cycle. The performance of these experiments is summarized in Table \ref{TAB:Sensitivities}.

\begin{table}
	\caption{\label{TAB:Sensitivities}A comparison of $\bar{d}$ flux sensitivity and kinetic energy capabilities for cosmic ray antimatter experiments. T$_{min}$ and T$_{max}$ refers to the minimum and maximum kinetic energy per nucleon, $\Phi_{min}$ refers to the minimum detectable flux. Values for GAPS are preliminary and subject to change due to ongoing detector design revisions.}
	\vspace{10pt}
	\begin{tabular}{llll}
		Experiment & T$_{min}$---T$_{max}$ [MeV/n] & $\Phi_{min}$ [((GeV/n) m$^2$ s sr)$^{-1}$] & Reference \\ \hline
		GAPS & 50-200 & $\approx 1.5\times10^{-6}$ & \cite{aramaki_16} \\
		BESS & 170-1155 & $1.8\times10^{-4}$ & \cite{bess_05} \\
		AMS-02 & 200-790, 2430-4200 & $2.1\times10^{-6}$ & \cite{ams02_08} \\ \hline
	\end{tabular}
\end{table} 

\section{The GAPS instrument}
\begin{figure}
	\begin{center}
	\includegraphics[scale=0.37]{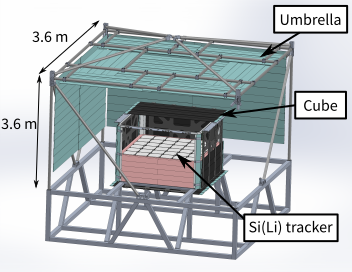}
	\end{center}
	\caption{\label{FIG:gondola}3D cutaway of the gondola assembly with main systems labeled.}
\end{figure}

While GAPS is eponymously referred to as a spectrometer, it is a very new take on the traditional design. The major distinction is the absence of a magnet. Instead, GAPS relies on tightly integrated time of flight and lithium drifted silicon, or Si(Li), tracker system. Eliminating the mass and volume requirements of the magnets presents numerous advantages, the most important being an increased active volume. A 3D view of the detector concept with important labeled systems is shown in Figure \ref{FIG:gondola}. In this section we describe the operating principles, general design features, expected sensitivities and the construction progress to date.

\subsection{Particle detection}
A cosmic ray primary enters the detector through two hermetic layers of a scintillator time of flight (TOF) system. While the main purpose of the TOF is particle $\beta$ measurement, it will be used to identify the primary species based on the amount of light generated ($\propto Z^2 / \beta^2$), as well as serving as the global trigger.

\begin{figure}
	\includegraphics[scale=1.3]{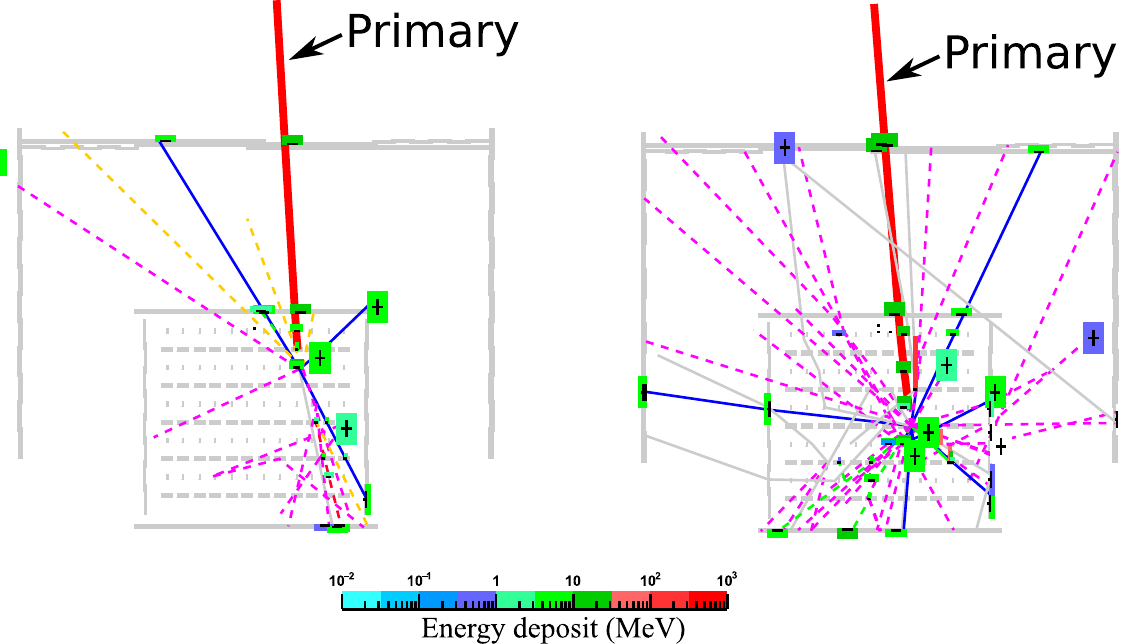}
	\caption{\label{FIG:det-concept}Event browser for simulations using a development version trigger algorithm and Geant4 detector mockup. Color convention (see electronic version): \textcolor{blue}{pion}, \textcolor{magenta}{photon}, \textcolor{green}{electron}(dashed), \textcolor{red}{muon}. This is a side view; the umbrella, inner cube and Si(Li) layer outlines are drawn.}
\end{figure}

The main detector target consists of multiple layers of Si(Li) wafers. As the primary transits through the TOF and Si(Li) layers it continuously loses kinetic energy until it eventually stops in the tracker volume---with $\bar{d}$ tending to stop deeper than $\bar{p}$. Since the primaries are cosmic ray antimatter replacing a shell $e^{-}$, they will briefly form exotic atoms which then de-excite resulting in the annihilation of the target nucleus. During de-excitation, atomic X-rays of specific energies are produced which depend on the captured primary mass. The annihilation event will produce a number of pions and protons. The multiplicity of these particles is combined with the X-ray information to discriminate the species of the primary. An example event is shown in Figure \ref{FIG:det-concept}. The qualitative difference between $\bar{p}$ ($T=130.2$ MeV) and $\bar{d}$ ($T/n=142.3$ MeV/n) is apparent: the pion multiplicity for the $\bar{p}$ event is about half the $\bar{d}$ event. There are also more secondaries in general for the $\bar{d}$ event, such as electrons and protons. For these examples, $\bar{d}$ results in 24 TOF hits and 22 Si(Li) tracker hits, while $\bar{p}$ gives 12 TOF hits and 9 Si(Li) tracker hits.

\newpage

\subsection{Design and layout}
\begin{figure}
	\begin{center}
	\includegraphics[scale=0.72]{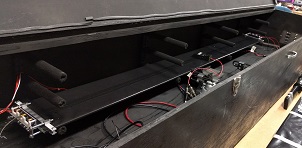}
	\includegraphics[scale=1]{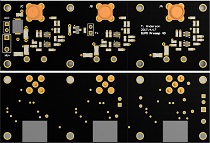}
	\end{center}
	\caption{\label{FIG:V2_paddle}\textit{Left:} Umbrella type paddle with V2 preamp and SiPM coupled to each end. \textit{Top right:} reverse side of V2 preamp with connectors and components. \textit{Bottom right:} front side of V2 preamp with SiPMs attached (gray squares).}
\end{figure}

The apparatus includes four major systems
\begin{enumerate}
	\item TOF
	\item Si(Li) tracker
	\item Flight computer, NASA managed communications, power distribution
	\item Oscillating heat pipe (OHP) cooling
\end{enumerate}
which are designed to be mounted on a typical long duration balloon (LDB) gondola. The TOF active medium is PVT scintillator (Eljen-200) and consists of two main components: an outer ``umbrella'' section, and inner ``cube'' section. The umbrella segments use 180 cm $\times$ 16 cm $\times$ 0.635 cm paddles. The cube will use  156(106.5) cm $\times$ 16(10) cm $\times$ 0.635 cm side (corner) paddles. Near UV total internally reflected scintillation photons will be observed using silicon photomultipliers ($6\times$ Hamamatsu S14160-6050HS) optically coupled to the paddle ends. All sensors are combined into a dual-stage resistively loaded voltage amplifier (RLVA), which branches to three outputs: low gain (equivalent to one SiPM output), high gain (tuned for a specific dynamic range), and an experimental shaped high gain channel (for timing studies). A photo of an umbrella paddle and the current preamp revision (V2) are shown in Figure \ref{FIG:V2_paddle}.

\begin{figure}
	\begin{center}
	\includegraphics[scale=0.5]{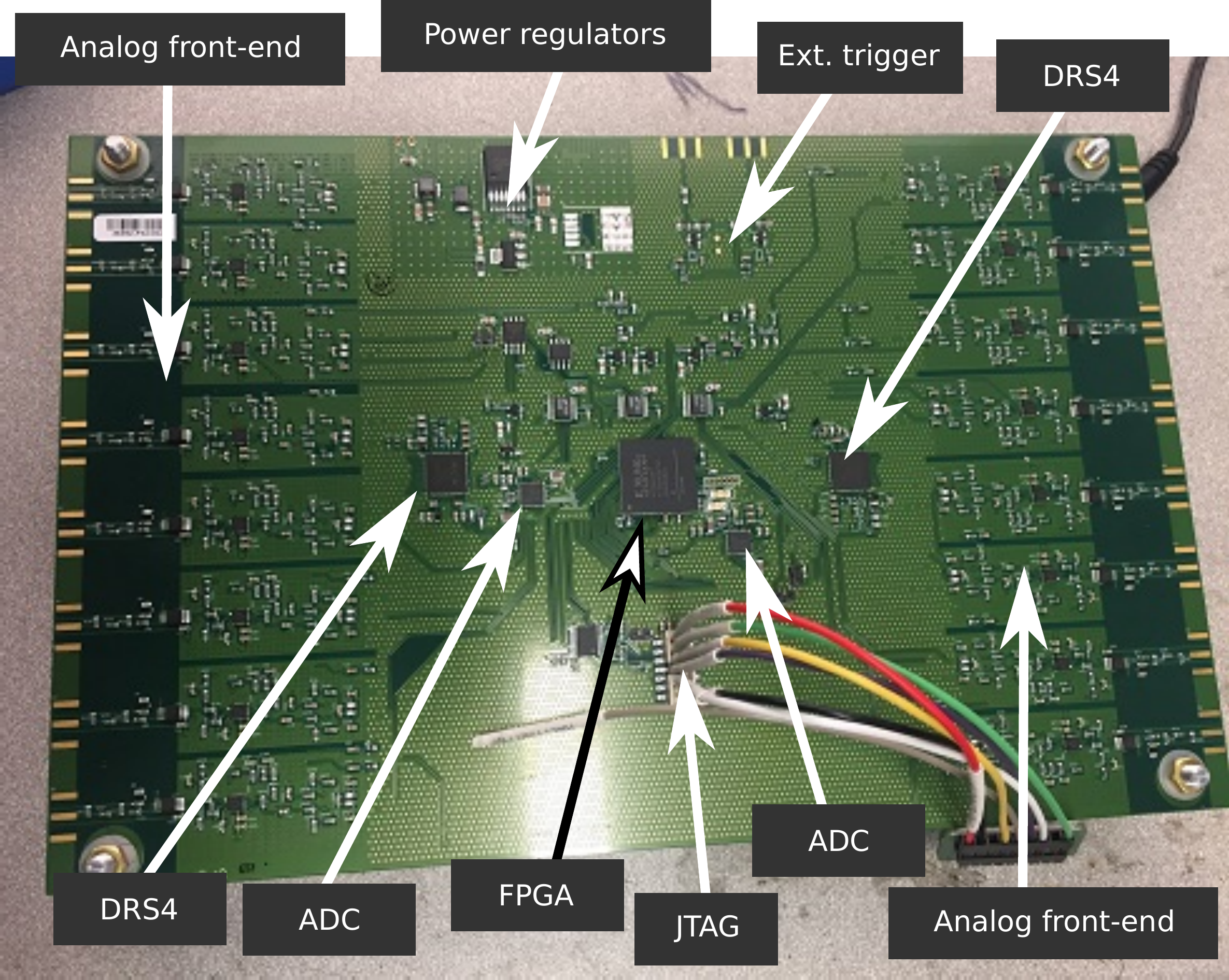}
	\end{center}
	\caption{\label{FIG:readout}First revision of UCLA TOF readout board, received early April 2018.}
\end{figure}

The high gain output from 8 paddles (for 16 total channels) will be routed to a readout board for digitization. To achieve sub-nanosecond accuracy while satisfying power and cost requirements, the DRS4 switched capacitor array ASIC (application specific integrated circuit) \cite{ritt_04} will be used. The current readout board consists of an analog front-end, $2\times$ DRS4, $2\times$ ADC and FPGA which controls the DRS4 and also implements an SPI communications system. A photo showing the main components is provided in Figure \ref{FIG:readout}. The next revision will move away from a monolithic design, favoring a mezzanine approach with 3 layers.

In addition to the readout board, the analog signals are also split and connected to a local trigger board which implements various discrimination levels and combinatorial logic to search for events. These boards are connected to a master trigger system which makes a system-wide trigger decision based on local patterns. The master trigger will initiate waveform readout on the TOF and Si(Li) system.

\begin{figure}
	\begin{center}
	\includegraphics[scale=0.185]{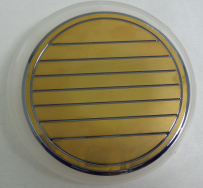}
	\includegraphics[scale=1.5]{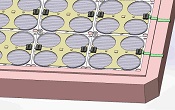}
	\end{center}
	\caption{\label{FIG:sili}\textit{Left:} 8-strip version of the Si(Li) detector manufactured by Shimadzu Corp. \textit{Right:} Conceptual schematic of section of Si(Li) layer, with detectors mounted to electronics board.}
\end{figure}

The Si(Li) tracker system will consist of 10 layers partitioned into modules populated with 4 Si(Li) wafers (1440 total). For an 8-strip detector this results in 11,5200 total input channels. Signals are buffered with a charge sensitive amplifier (CSA) and sent into a 32-channel custom ASIC for digitization and readout. The detectors provide optimal X-ray energy resolution at colder temperatures and higher bias, and will nominally be operated around -40$^{\circ}$ C and 200 V. A custom ASIC is also being developed for the high voltage power supply and data acquisition back-end system which is directly interfaced to the flight computer. Figure \ref{FIG:sili} shows a recent detector prototype and diagram of layer assembly.

The cooling system will use a passive oscillating heat pipe (OHP) design which reduces mass and power requirements while also saving on cost. The cooling mechanism relies on liquid cylinders separated by vapor bubbles. When heat is applied to the evaporator (from electronics heat sink, for example) the vapor bubbles expand, pushing the fluid toward a condenser, where the heated gas condenses back to liquid phase. This generates oscillating action circulating the fluid through the capillary tubes.

\subsection{Sensitivity and expected performance}
For an altitude of 36 km (roughly 5 g cm$^{-2}$ overburden) and $3\times35$ long duration balloon flights, the expected minimum $\bar{d}$ sensitivity is $\approx 1.5\times10^{-6}$ ((GeV/n) m$^2$  s sr)$^{-1}$, and the expected minimum $\bar{p}$ sensitivity is $\approx 2\times10^{-3}$ ((GeV/n) m$^2$  s sr)$^{-1}$. While not optimized for $\bar{\mathrm{He}}$, GAPS will be able sensitive to this species and efforts are underway to understand its signature.

\begin{figure}
	\begin{center}
	\includegraphics[scale=0.45]{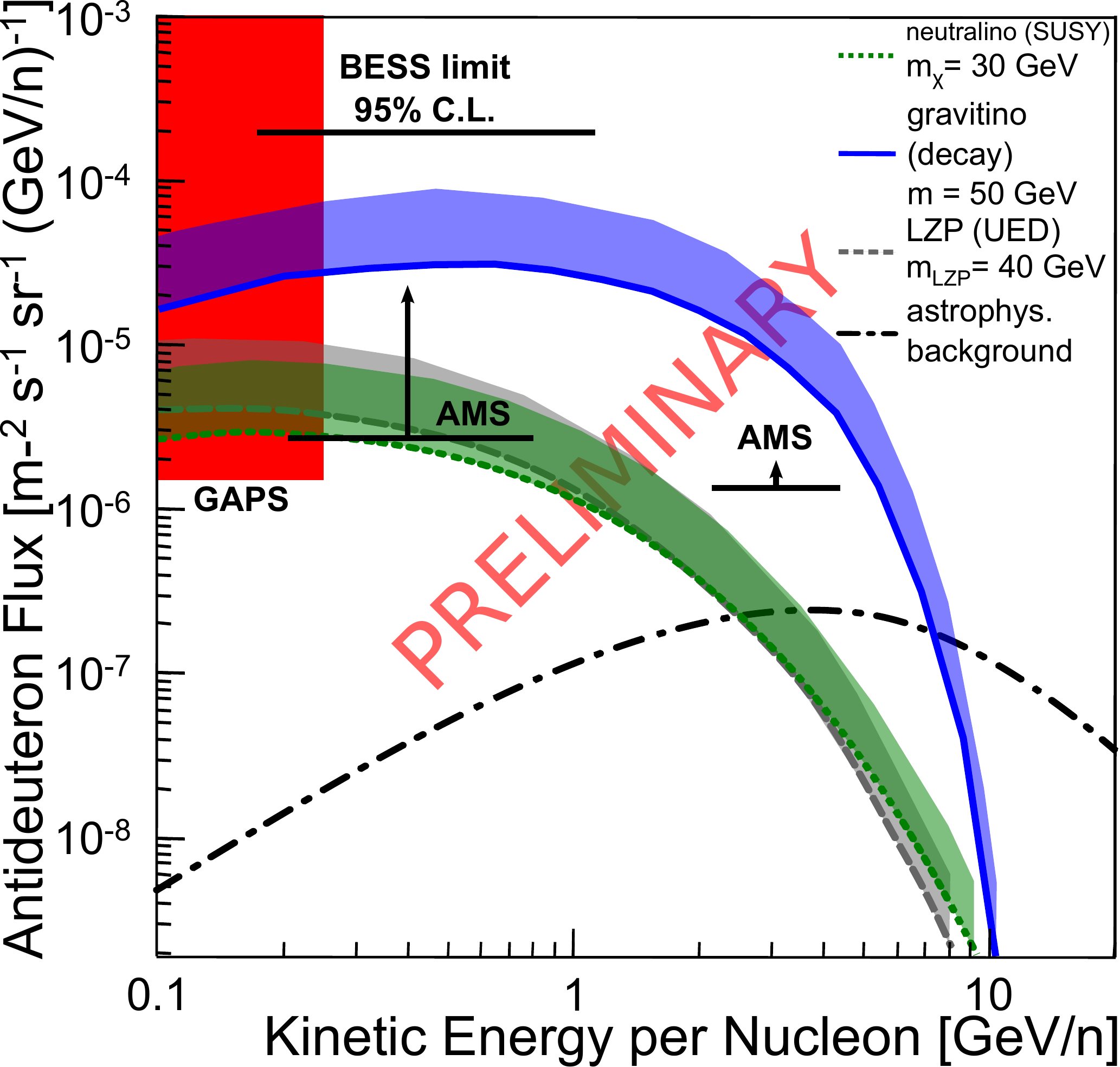}
	\end{center}
	\caption{\label{FIG:dbar_sens}Projected flux sensitivity for GAPS and associated antimatter experiments. Three leading BSM predictions are also shown, along with model uncertainty due to propagation.}
\end{figure}

In Figure \ref{FIG:dbar_sens} the $\bar{d}$ energy spectrum is shown along with a few beyond the standard model (BSM) predictions. It can be seen that GAPS has the potential to place strong constraints on low mass WIMPs, low mass gravitinos from supergravity theories, and low mass Kaluza-Klein particles predicted by extra-dimensional theories. The primary to secondary flux ratio is $\approx50$ at 200 MeV/$n$ energies and up to $\approx300$ toward lower energies.

\subsection{Construction progress}
To capitalize on the expected minimum of solar activity (end of cycle 24) in the austral summer of 2019-2020, GAPS is proceeding with a rapid development timeline. The first revision of the TOF readout board has been tested with satisfactory performance and will be used to construct a TOF engineering model in the fall of 2018. The second revision is based on an updated architecture which moves to a mezzanine design: a top level logic board based on the Zynq-7000 system on module, a second level break out board which includes Ethernet communications, and a third level mixed signal board which includes the analog front-end, DRS4 and FADCs.

Recently the first revision of the digitizing ASIC for the Si(Li) tracker was submitted for fabrication. A prototype electronics board has also been assembled and is currently being tested. Collaborators working closely with Shimadzu Corp.~have tested and characterized a number of Si(Li) prototypes demonstrating 3.8 keV energy resolution for 59.5 keV X-rays from a $^{241}$Am source.

Significant progress has been made in the mechanical design of both the Si(Li) and TOF superstructure. Prototype schematics and assemblies are nearing completion and several mock components for the TOF system have been fabricated and tested. Extensive simulation work is ongoing for the cooling system and a test of a scaled down prototype has been scheduled.

\section{Conclusion}
The GAPS experiment will use a novel detection technique to expand the antiproton energy spectrum to lower energies, and provide independent but complementary measurements to BESS and AMS-02 at higher energies. The design is capable of measuring low energy $\bar{d}$ with very high sensitivities. The primary motivation for both measurements is indirect dark matter search for low mass WIMPs. Additionally, GAPS will provide potentially useful constraints for the modeling of low energy galactic cosmic ray propagation, and other astrophysical applications. Early revisions of key hardware systems have recently been assembled and tested, and future revisions are being developed. GAPS is scheduled to launch in Antarctica in late 2020.

\Acknowledgements
SQ wishes to thank the organizers of CIPANP18 for the opportunity to share recent progress on the GAPS project. This work was supported, in part, by NASA grant NNX17AB45G.


\begin{thebibliography}{99}

\bibitem{Planck2015}
\href{https://doi.org/10.1051/0004-6361/201525830}{Planck 2015 results - XIII. Cosmological parameters}, Planck Collaboration, A\&A, 594 (2016) A13

\bibitem{Rubin1970}
\href{https://doi.org/10.1086/150317}{Rotation of the Andromeda Nebula from a Spectroscopic Survey of Emission Regions}, ApJ, 159 (1970) 379

\bibitem{Clowe2004}
\href{https://doi.org/10.1086/508162}{A Direct Empirical Proof of the Existence of Dark Matter}, Clowe et al., ApJL, 648 (2006) L109

\bibitem{OMeara2001}
\href{https://doi.org/10.1086/320579}{The Deuterium to Hydrogen Abundance Ratio toward a Fourth QSO: HS 0105+1619}, O'Meara et al., ApJ, 552 (2001) 718--730

\bibitem{Brandt2016}
\href{http://dx.doi.org/10.3847/2041-8205/824/2/L31}{Constraints on Macho Dark Matter From Compact Stellar Systems
	in Ultra-Faint Dwarf Galaxies}, T. Brandt, ApJL, 824 (2016) L31 

\bibitem{DMSurvey2018}
\href{https://doi.org/10.1088/1361-6633/aab913}{WIMP dark matter candidates and searches—current status and future prospects},  L.  Roszkowski, E. Maria Sessolo, and S. Trojanowski,  Rep. Prog. Phys. 81 (2018) 066201

\bibitem{Donato2018}
\href{https://doi.org/10.1103/PhysRevD.97.103011}{Prospects to verify a possible dark matter hint in cosmic antiprotons with antideuterons and antihelium}, M. Korsmeier, F. Donato, and N. Fornengo, Phys. Rev. D 97 (2018) 103011

\bibitem{Donato2008}
\href{https://doi.org/10.1103/PhysRevD.78.043506}{Antideuteron fluxes from dark matter annihilation in diffusion models}, F. Donato, N. Fornengo, and D. Maurin, Phys. Rev. D 78 (2008) 043506

\bibitem{Boezio_97}
\href{http://dx.doi.org/10.1086/304593}{The Cosmic-Ray Antiproton Flux between 0.62 and 3.19 GeV Measured Near Solar Minimum Activity}, M. Boezio et al., ApJ 487 (1997) 415

\bibitem{Boezio_01}
\href{https://doi.org/10.1086/323366}{The Cosmic-Ray Antiproton Flux between 3 and 49 GeV}, M. Boezio et al., ApJ 561 (2001) 787

\bibitem{Asaoka_02}
\href{https://doi.org/10.1103/PhysRevLett.88.051101}{Measurements of Cosmic-Ray Low-Energy Antiproton and Proton Spectra in a Transient Period of Solar Field Reversal}, Y. Asaoka et al., Phys. Rev. Lett. 88 (2002) 051101

\bibitem{Abe_08}
\href{https://doi.org/10.1016/j.physletb.2008.10.053}{Measurement of the cosmic-ray low-energy antiproton spectrum with the first BESS-Polar Antarctic flight}, K. Abe et al., Phys. Lett. B 670 (2008) 103

\bibitem{Abe_17}
\href{https://doi.org/10.1016/j.asr.2016.11.004}{The results from BESS-Polar experiment}, Adv. in Space Res. 60 (2017) 806-814

\bibitem{ams_16}
\href{https://doi.org/10.1103/PhysRevLett.117.091103}{Antiproton Flux, Antiproton-to-Proton Flux Ratio, and Properties of Elementary Particle Fluxes in Primary Cosmic Rays Measured with the Alpha Magnetic Spectrometer on the International Space Station}, M. Aguilar et al. (AMS Collaboration), Phys. Rev. Lett. 117 (2016) 091103

\bibitem{aramaki_14}
\href{https://doi.org/10.1016/j.astropartphys.2014.03.011}{Potential for precision measurement of low-energy antiprotons with GAPS for dark matter and primordial black hole physics}, T. Aramaki et al., Astropart. Phys. 59 (2014) 12-17

\bibitem{aramaki_16}
\href{https://doi.org/10.1016/j.astropartphys.2015.09.001}{Antideuteron sensitivity for the GAPS experiment}, T. Aramaki et al., Astropart. Phys. 74 (2016) 6-13

\bibitem{bess_05}
\href{https://doi.org/10.1103/PhysRevLett.95.081101}{Search for Cosmic-Ray Antideuterons}, H. Fuke et al., Phys. Rev. Lett. 95 (2005) 081101

\bibitem{ams02_08}
\href{http://indico.nucleares.unam.mx/event/4/session/40/contribution/1112/material/paper/0.pdf}{Cosmic Rays Antideuteron Sensitivity for AMS-02 Experiment}, V. Choutko et al., Proc. of the 30th Intl Cosmic Ray Conf. 4 (2008) 765-768

\bibitem{ritt_04}
\href{https://doi.org/10.1016/j.nima.2003.11.059}{The DRS chip: cheap waveform digitizing in the GHz range}, S. Ritt, Nucl. Instrum. Meth. A 518 (2004) 470-471


\end{thebibliography}
\end{document}